\documentclass{sf2a-conf2014}
\usepackage{graphicx}
\usepackage{hyperref}
\usepackage{amsmath,amsthm, amssymb, latexsym}
\usepackage[]{natbib}  
\usepackage{epstopdf}

\def\BibTeX{{\rm B\kern-.05em{\sc i\kern-.025em b}\kern-.08em
    T\kern-.1667em\lower.7ex\hbox{E}\kern-.125emX}}
\bibpunct{(}{)}{;}{a}{}{,}  %%%%%%%%%%%%%  A&A bibliography style
\begin{document}

\TitreGlobal{SF2A 2014}

%%-----------------------------------------------------------------
%%      the top matter
%%

\title{Automated procedure to derive fundamental parameters of B and A stars: Application to the young cluster NGC 3293}

\runningtitle{Automated procedure to derive fundamental parameters}

\author{E. Aydi}\address{Department of Physics \& Astronomy, Notre Dame Univeristy - Louaize, Zouk Mikael Lebanon}

\author{M. Gebran$^1$}

\author{R. Monier$^{2,}$}\address{LESIA/CNRS UMR 8109, Observatoire de Paris -- Universit\'e Pierre et Marie Curie -- Universit\'e Paris Denis Diderot, 5 place Jules Janssen, 92190 Meudon, France}\address{Laboratoire Lagrange, Universit\'e de Nice Sophia Antipolis, Parc Valrose, 06100 Nice, France}

%% IF Author3 has the same affiliation than Author1:
%%\author{E. Author3$^1$}

%% IF Author3 has its own affiliation:
\author{F. Royer}\address{GEPI/CNRS UMR 8111, Observatoire de Paris -- Universit\'e Paris Denis Diderot, 5 place Jules Janssen, 92190 Meudon, France}

%% IF Author3 has two affiliations, the one of Author1 and a second one:
%%\author{C.\,E. Author3$^{1,}$}\address{Dept. of Chess, University of Games, 35101 Las Vegas, Monaco} 

\author{A. Lobel}\address{Royal Observatory of Belgium, Ringlaan 3, B--1180 Brussels, Belgium}
\author{R. Blomme$^5$}

%% Keep this line, even if the page will be settled afterwards.
\setcounter{page}{237}

%%-----------------------------------------------------------------

\maketitle

%%-----------------------------------------------------------------
%%        The abstract
%% 
%%  Warning!  within the abstract:
%%  - do not use macros. 
%%  - do not use commands like: \cite, \citet, \$\xi_t$citep ... etc.

\begin{abstract}
This work describes a procedure to derive several fundamental parameters such as the effective temperature, surface gravity, equatorial rotational velocity and microturbulent velocity.  In this work, we have written a numerical procedure in Python which finds the best fit between a grid of synthetic spectra and the observed spectra by minimizing a standard chi-square. LTE model atmospheres were calculated using the ATLAS9 code and were used as inputs to the spectrum synthesis code  SYNSPEC48 in order to compute a large grid of synthetic Balmer line profiles. This new procedure has been applied to a large number of new observations (GIRAFFE spectra) of B and A  stars members of the young open cluster NGC3293. These observations are part of the GAIA ESO Survey. Takeda's procedure was also used to derive rotational velocities and microturbulent velocities. The results have been compared to previous determinations by other authors and are found to agree with them. As a first result, we concluded that using this procedure, an accuracy of $\pm$ 200 K could be achieved in effective temperature and $\pm$ 0.2 dex in surface gravities.
\end{abstract}

%% Insert the keywords (to appear in the ADS indexing)
%% Keywords must be separated by a comma

%%-----------------------------------------------------------------

\section{Introduction}
Open clusters are of tremendous importance to astrophysics. Stars in clusters serve as "laboratories" for astronomers, as all are nearly at the same distance and have the same age and initial chemical composition. Hence open clusters are key objects in the study of stellar evolution \citep{Edvardson93}. The determination of the effective temperature and surface gravity of stars in open clusters are instrumental in  assigning proper spectral types to each star and pre-requisites to the determination of other parameters as the projected rotational velocity, microturbulent velocity and chemical abundances \citep{Evans05}.
The quantity of available astronomical data has increased considerably with the implementation of extensive surveys such as SDSS and RAVE, and will continue to increase in the near future especially with the European Space Agency Gaia mission. Therefore, there is a need to analyze these data in an homogeneous and efficient way.\\
Gaia's Radial Velocity Spectrograph (RVS) will collect millions of stellar spectra at a resolution of R=11500. These spectra will allow us to derive fundamental parameters of these stars and the abundances of a few chemical elements.\\ 
The need of an automated procedure for classification of stars has been discussed recently using different codes and mathematical approaches. As an example, one can mention the MATISSE algorithm \citep{Recio-Blanco06} or \citet{Hekker09} semi-automated procedures.\\
In this preliminary work, we present the first steps of an automated procedure aiming at determining several stellar parameters. The procedure is applied to a large sample of A and B stars members of the young open cluster NGC 3293 with an age $\sim$ 10 Myrs. The observations are part of the Gaia-ESO public survey. 
The effective temperatures ($T_{\rm{eff}}$) and surface gravities ($\log g$) of the selected stars were determined by adjusting a large grid of synthetic spectra to the observed ones, in particular to The Balmer line profiles. LTE model atmospheres were calculated using the ATLAS9 code \citep{Kurucz92} and were used in order to compute a large grid of synthetic Balmer line profiles using SYNSPEC48,  \cite{Hubeny92} spectrum synthesis code. \cite{Takeda95} procedure was used in order to determine the projected rotational velocity $v_e \sin i$ and the microturbulent velocity$\xi_t$.
  
\section{Observation and data reduction}
The selected stars are members of the young open cluster NGC 3293 which was observed in the frame of the Gaia-ESO survey. These spectroscopic data consist of three spectral ranges, two of which sample H$_{\delta}$ [4030 $\AA$ - 4200 $\AA$] and H$_{\alpha}$ [6300 $\AA$ - 6500 $\AA$], the third one samples several lines of the iron-peak elements from 4350 $\AA$ up to 4550 $\AA$.
The targets were observed using the FLAMES-GIRAFFE spectrograph with a resolution of R $\sim$ 25000 mounted on the unit 2 telescope (UT2) of the European Southern Observatory (Paranal-Chile).\\
The selected spectra are calibrated using GIRAFFE ESO pipeline which involves mainly bias subtraction, scattered light removal, bad pixel masking, flat-fielding, extraction, and wavelength calibration.\\
After this reduction the flux is given in arbitrary units. The normalization of the spectrum is done using the IRAF (Image Reduction and Analysis Facility) software \citep{Tody86}. To ensure we correctly located regions free of lines (when available) in each order, we have computed synthetic spectra using the code SYNSPEC48 \citep{Hubeny92} assuming a solar metallicity for the various temperatures and surface gravities of our stars. The spectra were then rectified to the local continuum. 
%\begin{figure}[ht!]
% \centering
% \includegraphics[width=0.8\textwidth,clip]{Aydi_fig1}      
%  \caption{\textbf{Left}: Observed spectrum im arbitrary flux. \textbf{Right}: The same spectrum, normalized to the continuum. Wavelenghts are in Angstroms.}
%  \label{Aydi:fig1}
%\end{figure}

\section{Model atmospheres and synthetic spectra calculations}

The calculations of the LTE model atmospheres were carried out using Kurucz's ATLAS 9 code \citep{Kurucz92}. Model atmospheres were computed assuming a plane parallel geometry, hydrostatic equilibrium, radiative equilibrium and depth independent turbulent velocity fixed to 2 km.s$^{-1}$ and solar abundances \citep{Grevesse98}. We ran ATLAS 9 for effective temperatures ranging from 6000 K up to13000 K and for surface gravities from 3.0 dex up to 5.0 dex.

The model atmospheres computed with ATLAS 9 served as inputs to SYNSPEC48.
Two important files are read as inputs before the computation of the theoretical flux: the ATLAS9 model atmosphere and the linelist for each of the three spectral regions aforementioned (compiled from \cite{Gebran10}). Balmer lines were calculated using these set of model atmospheres for different $T_{\rm{eff}}$ and $\log g$ and concolved with i) a Gaussian profile having the instrumental FWHM and ii) a parabolic rotational profile
 for the appropriate $v_e \sin i$. All convolved spectra have a resolving power R $\sim$ 25000.\\
The grid obtained from SYNSPEC48 consists of $\sim$ 230000 synthetic spectra whose effective temperatures range from 6000 K to 13000 K calculated with a step $\Delta(T_{\rm{eff}})$ = 50 K, the surface gravities range from 3.0 dex to 5.0 dex calculated with a step $\Delta(\log g)$ = 0.05 dex, and the projected rotaional velocities range from 0 to 200 km/s with a step $\Delta(v_e \sin i)$ = 5 km/s.

\section{The procedure}

%\subsection{Python Programming}

Our procedure, written in Python, looks for the best fit between a grid of synthetic spectra and the observed one. It performs the following sequence:\\
- Reading the data : the procedure reads both the synthetic and  the observed spectra.\\
- Interpolation : An interpolation is performed so that the theoretical fluxes be evaluated at exactly the same wavelengths as the observed spectra.\\
- Radial velocity and scaling factors: radial velocity is derived during the procedure. A scaling factor was introduced so as to correct for any residual slope after continuum normalization.\\
- Chi-square calculation and minimization between observed and synthetic spectra from the grid.\\
- The results :  $T_{\rm{eff}}$, $\log g$, $v_e \sin i$, and the radial velocity are derived automatically after running the procedure using our synthetic grid.
All results were checked visually by over-plotting the best fit to the observation.\\

Tests were carried out over the whole range of $T_{\rm{eff}}$ and $\log g$ in order to check the efficiency of the procedure. As a test, we chose one of the synthetic spectrum and used it as a surrogate observed spectrum. After including all kinds of effects such as radial velocity shift, scaling factor, a decrease in the S/N, we always ended up having a consistent result.\\
In order to minimize the calculation time per observation,  we attempted to decrease the number of computed spectra in our grid (originally $\sim$ 230,000) leading to an increase of the fundamental parameters steps. For that reason, we have specifically studied the effect of each parameter alone on the final result over the entire range of effective temperatures, surface gravities and rotational velocities, our tests show that adopting a step of 100 K in $T_{\rm{eff}}$, 0.1 dex in $\log g$, and 50 km/s in $v_e \sin i$ leads to an accuracy of of $\pm$ 200 K in effective temperature and $\pm$ 0.2 dex in surface gravities. Using these steps will reduce the size of the grid to $\sim$ 7000 synthetic spectra which shortens the calculation time for each observation from 5 days to $\sim$ 4 hours.

\section{Determination of $v_e \sin i$ and $\xi_t$}

The second part of this work consisted in deriving accurate values for $v_e\sin i$ and $\xi_t$. These two parameters were derived iteratively using \cite{Takeda95} procedure which minimizes the chi-square between the normalized synthetic spectrum and the observed one. The synthetic spectra were calculated using the previously determined $T_{\rm{eff}}$ and $\log g$.\\
Technically, rotational velocities ($v_{e}\sin i$) and microturbulent ($\xi_{t}$) velocities were derived using several weak and moderately strong FeII  lines located between 4491.405\,\AA \ and 4508.288\,\AA \ and the~MgII triplet at 4480\,\AA\ and assuming solar abundances.

\section{Results}

\begin{figure}[ht!]
 \centering
 \includegraphics[scale=0.5]{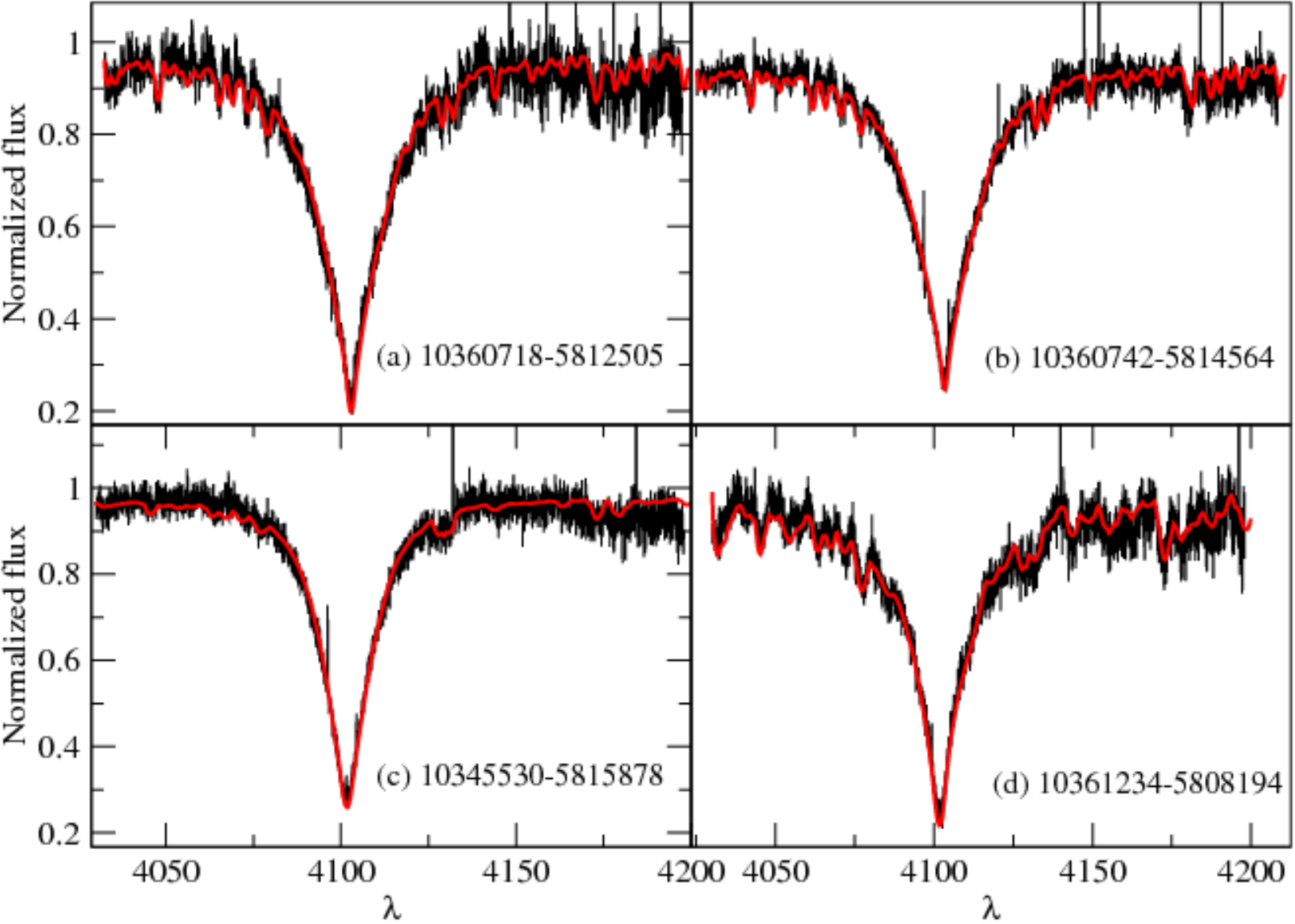}      
 \caption{Synthetic spectra (in red) superimposed to the observed ones (in black) for 4 stars member of NGC3293.}
 \label{Aydi:fig1}
\end{figure}

The procedure was applied to 80 stars. $T_{\rm{eff}}$, $\log g$, and preliminary values of $v_e\sin i$ was derived for all stars. More accurate values of $v_e\sin i$ and $\xi_t$ were derived for 32 stars using \cite{Takeda95} iterative procedure. Fig.~\ref{Aydi:fig1} displays an example of the best fit between observed spectra (in black) and the synthetic ones (in red).\\

\begin{figure}[ht!]
 \centering
 \includegraphics[scale=0.3]{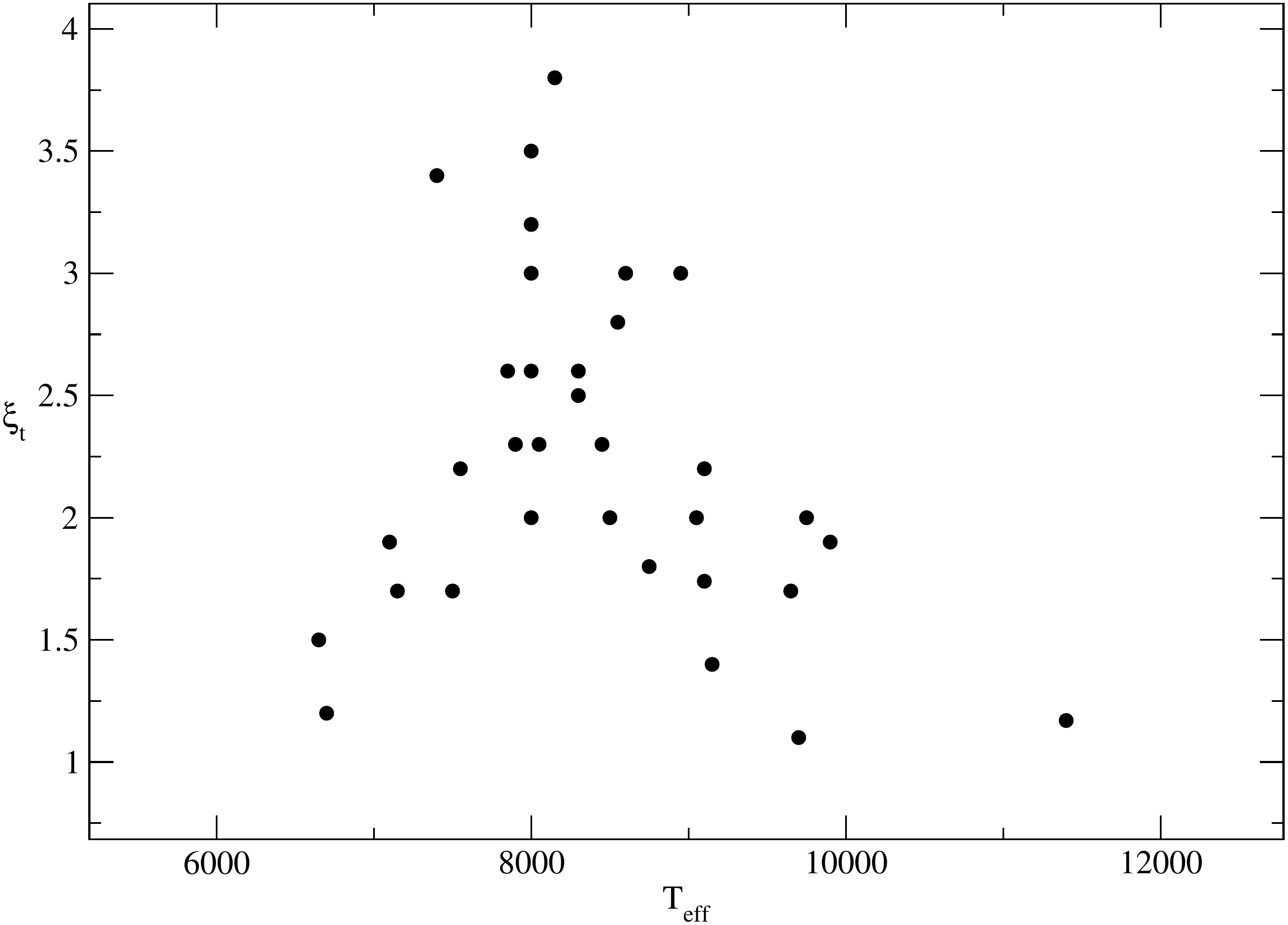}  
  \includegraphics[scale=0.3]{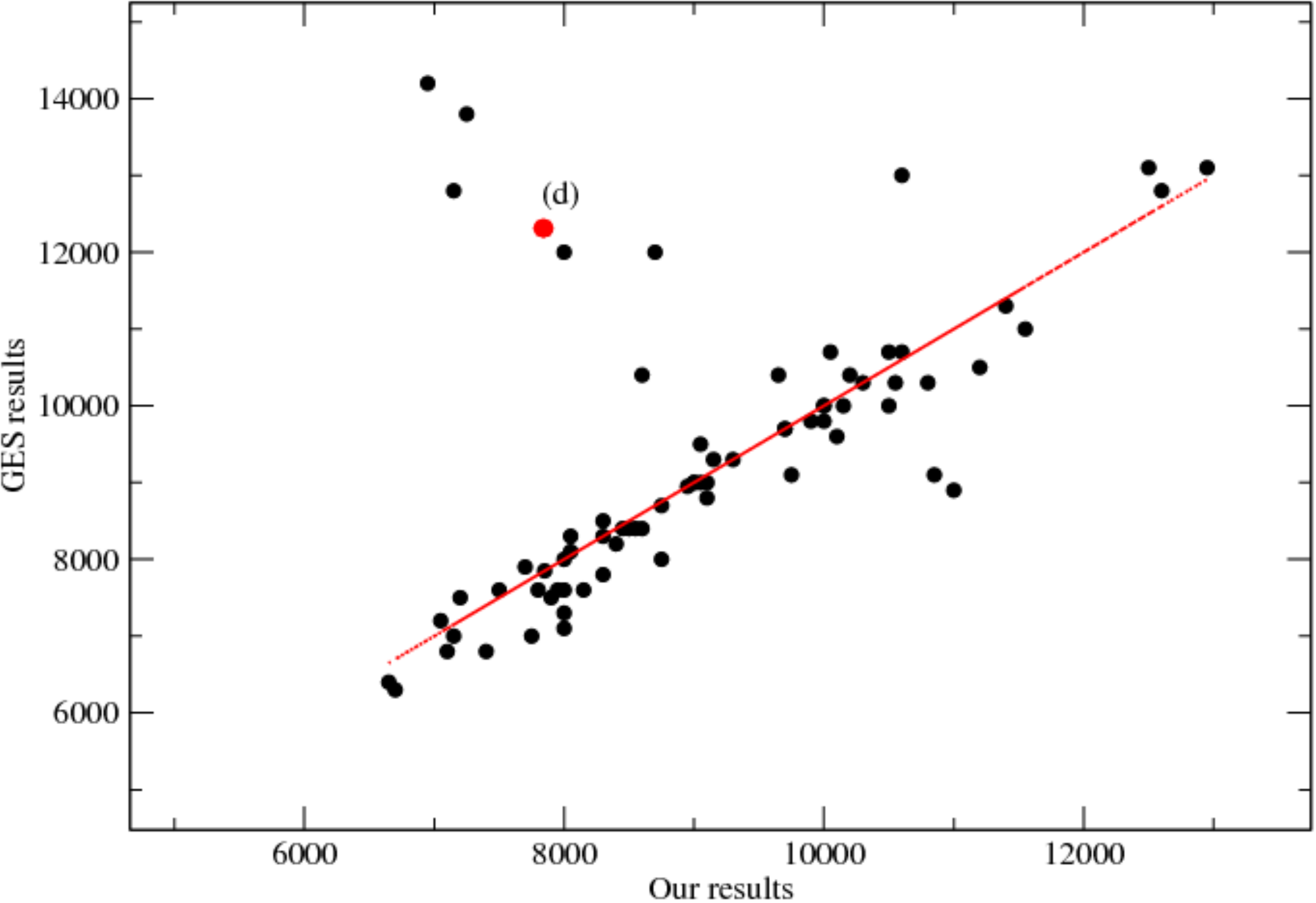}      
     
 \caption{\textbf{Left}: Microturbulent velocity in km/s as a function of effective temperature $T_{\rm{eff}}$ in K for the selected stars. \textbf{Right}: Comparison between the derived effective temperatures and the ones from the CASU GES archive (v2.1), as calculated by S. Koposov. }
 \label{Aydi:fig2}
\end{figure}

The left part of Fig.~\ref{Aydi:fig2} displays the derived microturbulent velocities as a function of the effective temperatures derived from Takeda's procedure. These results show that $\xi_t$ reaches its maximum around 8000 K (3$\pm$1 km/s) then decreases to 1 km/s around 6000 K and for high temperatures. This result agrees well with previous determinations : \cite{Gebran07}, \cite{Gebran13}, \cite{Takeda08} and with \cite{Smalley04} prescriptions for convection for tepid stars.\\

The right part of Fig.~\ref{Aydi:fig2} displays the comparison between our derived effective temperature and the ones from the CASU GES archive (v2.1), as calculated by S. Koposov. We found very good agreement for most of the stars. Large discrepabcies exist for a few stars only which are assigned large temperatures in the GES archive. Our determinations differ by up to $\sim$6000 K which is far larger than the expected error bar. The symbol depicted in red in the right part of Fig.~\ref{Aydi:fig2} corresponds to the observed spectra displayed in Fig.~\ref{Aydi:fig1}(d). Our derived temperature for this star is about 4000 K lower than the one derived by the GES team whereas the surface gravities are very close.

\section{Conclusions and future works}
The present procedure provides a fast automatic tool to provide estimates of the effective temperatures and surface gravities for a large number of stars. Hence, this procedure can be applied to large spectroscopic surveys (RAVE, Gaia, ...). The accuracy achieved using this procedure is about$\pm$ 200 K in $T_{\rm{eff}}$ and $\pm$ 0.2 dex  in $\log g$. The derived microturbulent velocities agree well with \cite{Gebran07}, \cite{Gebran13}, \cite{Takeda08} and with \cite{Smalley04} prescriptions for B, A, and F stars.\\

This procedure is currently being improved so that it can derive the projected rotational velocity and microturbulent velocity without having to resort to Takeda's procedure. The results should be constrained using another Balmer line such as the H$_{\alpha}$ profile. More refinements such as adopting depth-dependent abundances and microturbulent velocities should ultimately be brought to the procedure.
The derived parameters will ultimately be used to derive detailed elemental abundances for the studied stars and constrain physical processes at work in the radiative zones of tepid stars.

%The format for references is the one adopted by A\&A. To set the reference list in the proper format, we encourage you to use \BibTeX~ and the natbib package instead of the standard \verb=\thebibliography= environment.

% Optional acknowledgements
% -------------------------
\begin{acknowledgements}
Acknowledgements: This work is based on observations collected with the FLAMES spectrograph at the VLT/UT2 telescope (Paranal Observatory, ESO, Chile), for the Gaia-ESO Large Public Survey, programme 188.B-3002.
\end{acknowledgements}

%%-----------------------------
%%   Bibliography
%%-----------------------------
%%
%% The reference list should contain all the references cited in the text, ordered alphabetically by surname (with
%% initials following). If there are several references to the same first author, they should be entered according
%% to the following scheme:
%% 1. One author: chronologically
%% 2. Author, one co-author: alphabetically by co-author, then chronologically
%% 3. Author, two or more co-authors: chronologically.
%%
%% Please note that for papers that have more than five authors, only the first three should be given, followed
%% by "et al."
%%
%% The format for references is the one adopted by A&A (see the example below).
%%
%% To set the reference list in the proper A&A format, we encourage you to use BibTEX and the natbib
%% package instead of the standard 'thebibliography' environment.
%%

%% The following lines are required when using BibTEX (strongly encouraged!):
\bibliographystyle{aa}  % A&A bibliography style file (aa.bst)
\bibliography{Aydi} % your references in file: Yourfile.bib

\end{document}